\documentclass{ws-ijgmmp}

\begin{document}

\markboth{Authors' Names}
{Instructions for Typing Manuscripts (Paper's Title)}

%
\catchline{}{}{}{}{}
%

\title{
	MIMETIC COMPACT STARS
	}

\author{D. MOMENI}

\address{Department of Physics, College of Science, Sultan Qaboos University,\\
	P.O. Box 36, P.C. 123, Al-Khodh, Muscat, Sultanate of Oman\\davood@squ.edu.om} 

\author{P.H.R.S. MORAES}

\address{ITA - Instituto Tecnol\'ogico de Aeron\'autica \\ Departamento de F\'isica, 12228-900, S\~ao Jos\'e dos Campos, S\~ao Paulo, Brazil\\
moraes.phrs@gmail.com }

\author{H. GHOLIZADE}
\address{Department of Physics, \\Tampere University of Technology P.O.Box 692, FI-33101 Tampere, Finland}

\author{R. MYRZAKULOV}
\address{Eurasian International Center for Theoretical Physics \\
	and Department of General Theoretical Physics, \\
	Eurasian National University, Astana 010008, Kazakhstan} 

\maketitle

\begin{history}
\received{(Day Month Year)}
\revised{(Day Month Year)}
\end{history}

\begin{abstract}
Modified gravity models have been constantly proposed with the purpose of evading some standard gravity shortcomings. Recently proposed by A.H. Chamseddine and V. Mukhanov, the Mimetic Gravity arises as an optimistic alternative. Our purpose in this work is to derive Tolman-Oppenheimer-Volkoff equations and solutions for such a gravity theory. We solve them numerically for quark star and neutron star cases. The results are carefully discussed.
\end{abstract}


\section{Introduction}\label{sec:int}

Since the late 90's, it is known that our universe is undergoing a phase of accelerated expansion. The first evidence came from Supernovae Ia observations \cite{riess/1998,perlmutter/1999}. Thenceforth, evidences from baryon acoustic oscillations \cite{eisenstein/2005} and cosmic microwave background anisotropies \cite{hinshaw/2013} seem to strengthen this cosmological scenario. If one assumes General Relativity (GR) as the background theory of gravity for a cosmological model, there is a necessity of considering $\sim70\%$ of the universe is filled with some exotic fluid named dark energy, which would be the responsible for the cosmic acceleration. However such a universe set up, although provide good agreement with observations, fails in explaining or describing some issues, like hierarchy, coincidence and cosmological constant problems \cite{randall/1999,arkani-hamed/1998,chimento/2003,cai/2005,weinberg/1989,sahni/2002}, among others. Some of those issues may, in principle, be solved (or at least evaded) by assuming alternative models of gravity.

An important class of alternative models arise from the assumption of a generalization of the Ricci scalar dependence on Einstein-Hilbert action, from which gravity field equations are derived, namely $f(R)$ theories \cite{sotiriou/2010}-\cite{starobinsky/2007}. 

A generalization of $f(R)$ theories has been recently proposed by T. Harko et al. \cite{harko/2011}, for which the gravitational part of the action depends not only on a function of $R$ but also on a function of $T$, the trace of the energy-momentum tensor.

Those theories have been intensively tested in the cosmological level (check, for instance, \cite{bamba/2014}-\cite{moraes/2015}). However, alternative models of gravity should also be tested in the astrophysical level. In fact, strong gravitational fields found in relativistic stars could discriminate standard gravity from its generalizations.

In what concerns $f(R)$ theory astrophysical tests, in \cite{momeni/2015}, the authors have obtained nonlocal-$f(R)=R+\alpha R^{2}$ Tolman-Oppenheimer-Volkoff (TOV) equations, which were solved numerically using adaptive Gaussian quadrature.  A general $f(R)$ model with uniformly collapsing cloud of self-gravitating dust particles was studied in \cite{cembranos/2012}. Such an investigation could be used as a first approximation to the modification that stellar objects can suffer in modified theories of gravity. In \cite{kainulainen/2007}, the gravitational field of stars were studied both analytically and numerically and generalized TOV equations have been derived.


Recently, another kind of alternative gravity model has been proposed, for which the metric is not considered a fundamental quantity. Instead, it is taken as a function of an auxiliary metric $\tilde{g}_{\mu\nu}$ and a scalar field $\phi$, contemplating the so called Mimetic Gravity (MG) \cite{Chamseddine:2013kea}. Note that such a dependence of the metric makes the application of the variational principle in the model action to yield more general equations of motion (EoM) than that of purely Einsteinian relativity theory.

In \cite{chamseddine/2014}, it was shown that by introducing a potential for the mimetic non-dynamical scalar field, one can mimic nearly any gravitational properties of normal matter. MG has been extended to $f(R)$ gravity in \cite{nojiri/2014} while cosmological viable models derived from such an extension are presented in \cite{momeni/2015b}.

Despite the references above, due to its recent elaboration, MG still presents a lack of cosmological and astrophysical application. Our purpose in this work is to verify MG at an astrophysical level, by analysing compact objects. By compact object in astrophysics we mean a homogeneous distribution of ordinary (or exotic) matter with energy-momentum tensor in the form of a perfect non-viscous fluid. The key note to start studying the dynamics of such systems is to write the EoM for a spherically static metric in Schwarzschild-Droste coordinates, denoted by $x^\mu =(t,r,\theta,\varphi)$. By rewriting the EoM in terms of the hydrodynamic parameters, hydrostatic pressure $p$, energy density $\rho$, and mass $M$ (active mass inside the star), we will obtain the TOV equations. The matter inside the compact star is composed of nuclear matter, sometimes in the form of strange quark matter \cite{weber/2005,farhi/1984,steinheimer/2014}, with a density of order $ \rho\sim10^{15}gr/cm^{3}$. It is believed that this matter has ordinary statistical properties and can enter a series of second order phase transitions \cite{Gholizade:2009ac}. 

We will derive TOV equations in the MG framework and solve them for different equations of state (EoS). The paper is organized as follows. In Section \ref{sec:mg} we briefly describe MG. In Section \ref{sec:mcs}, a model of compact star in MG is presented from the development of the TOV equations. In Section \ref{sec:ns}, numerical solutions for TOV equations are obtained from both neutron star (NS) and quarks star (QS) EoS, while in Section \ref{sec:dis}, we discuss the results.

\section{Mimetic Gravity}\label{sec:mg}
Conformal invariance plays an essential role to build gauge theory for gravity. Although GR is not conformal invariant in its original form, but it is possible to make conformal invariant theories in GR as well as in modified gravity \cite{Phys.Rept. 509 (2011) 167}. It has been demonstrated that spacetime deformations can generat  gravitational field and it has been shown that new exact solutions of the Einstein field equations can be derived  using deformations. Depending on type of gravitational theory, we can study metric deformation or other fields. In a case, metric deformations are considerd as a potentially effective way to address MG, where the deformed metric simplifies to
a generalization of the conformal transformations $g_{\mu\nu}\to \Omega(x^{\mu})g_{\mu\nu}$ called disformal transformations \cite{Bekenstein:1992pj}:

\begin{eqnarray}
&&g_{\mu\nu}\to \Omega(\phi,X)g_{\mu\nu}+B(\phi)\nabla_{\mu}\phi\nabla_{\nu}\phi,\ \ X:=\frac{1}{2}\nabla_{\mu}\phi\nabla_{\nu}\phi.
\end{eqnarray}
Here, $\phi$ is the scalar field and $X$ the kinetic term. if $\Omega(\phi,X)=1$, the deformation of a metric consists in adding to the background metric a new tensor \cite{Phys.Rept. 509 (2011) 167}.
A special case happens when we consider $B=0$ and $A=A(X)$, which is called MG. It was proven very recently that it can be interpreted as the same spirit as the disformal theory \cite{Deruelle:2014zza}.

In MG we  consider the physical metric $g_{\mu\nu}$ as a function of an auxiliary metric $\tilde{g}_{\mu\nu}$ and of a scalar field $\phi$. Therefore one has $g_{\mu\nu}=(\tilde{g}^{\alpha\beta}\partial_\alpha\phi\partial_\beta\phi)\tilde{g}_{\mu\nu}$. The MG action is therefore

\begin{equation}\label{eqn:mg2}
S=-\frac{1}{16\pi G_N}\int d^4x\sqrt{-g(\tilde{g}_{\mu\nu,\phi})}\{R[g_{\mu\nu}(\tilde{g}_{\mu\nu},\phi)]+\mathcal{L}_m\},
\end{equation} 
for which $\mathcal{L}_m$ is the matter lagrangian and $G_N$ denotes the Newtonian gravitational constant. Throughout this work we are going to assume units such that speed of light is  $c=1$.
By varying this action with respect to $\tilde{g}_{\mu\nu}$ and $\phi$ yields, respectively, the following EoM \cite{Chamseddine:2013kea}:

\begin{equation}\label{eqn:mg3}
(G^{\mu\nu}-8\pi G_N T^{\mu\nu})-(G-8\pi G_NT)g^{\mu\alpha}
g^{\nu\beta}\partial_\alpha\phi\partial_\beta\phi=0,
\end{equation}
\begin{equation}\label{eqn:mg4}
\frac{1}{\sqrt{-g}}\partial_\kappa[\sqrt{-g}(G-8\pi G_NT)g^{\kappa\lambda}\partial_\lambda\phi]=0,
\end{equation}
with $G$ and $T$ being the trace of the Einstein's tensor and of the energy-momentum tensor, respectively. The trace of the Equation (\ref{eqn:mg3}) is:
\begin{eqnarray}
&&(1-g^{\alpha\beta}\partial_\alpha\phi\partial_\beta\phi)(G-8\pi G_NT)=0.
\end{eqnarray}
In order to obtain a scenario dynamically different from GR, we suppose that $(G-8\pi G_NT)\neq0$, and consequently obtain the following equivalent EoM for the scalar field:
\begin{eqnarray}
&&g^{\alpha\beta}\partial_\alpha\phi\partial_\beta\phi=1.
\end{eqnarray}
This equation is in the Hamilton-Jacobi form and its general solution can be obtained by the same method. Generally speaking, $\phi$ can be a complex variable, i.e., $\phi\in\mathcal{C}$. 

It was shown in \cite{Chamseddine:2013kea} that such an approach could explain the cold dark matter. Here, instead, we will analyze the interior of compact objects from the MG TOV equations, which will be derived in the next section.  

\section{Model of a compact star in Mimetic Gravity}\label{sec:mcs}

We suppose that the metric of the compact star is static and spherically symmetric, with Schwarzschild-Droste coordinates $x^{\mu}=(t,r,\theta,\varphi)$ in the following form:
\begin{eqnarray}\label{eqn:mcs1}
ds^2= e^{2\psi}dt^2-e^{2\lambda}dr^2-r^2(d\theta^2+\sin^2\theta d\varphi^2),
\end{eqnarray}
for which the coefficients $\psi$ and $\lambda$ depend on the coordinate $r$ only. For such a metric, the non vanishing components of the Einstein's tensor read:
\begin{eqnarray}
G_{11}&=&-\frac{e^{2(\psi-\lambda)}}{r}\left(2\lambda'+\frac{e^{2\lambda}-1}{r}\right),\\
G_{22}&=&-\frac{1}{r}\left(2\psi' -\frac{e^{2\lambda}-1}{r}\right),\\
G_{33}&=&-re^{-2\lambda}[\psi'-\lambda'+r(\psi''+\psi'^2)-r\psi'\lambda'],\\
G_{44}&=&-\sin^2\theta G_{33},
\end{eqnarray}
for which primes denote derivation with respect to $r$. 

By considering the energy-momentum tensor of a perfect fluid, i.e., $T_{\mu}^{\nu}=diag(\rho,-p,-p,-p)$, with $\rho$ and $p$ representing the density and pressure of the star, respectively, the diagonal components of (\ref{eqn:mg3}) are:

\begin{eqnarray}\label{tt}
&&\frac{2\lambda'}{r}+\frac{e^{2\lambda}-1}{r^2}=8\pi G_N\rho  e^{2\lambda},
\end{eqnarray}
\begin{eqnarray}\label{rr}
&&\frac{2\psi'}{r}+\frac{1-e^{2\lambda}}{r^2}=-8\pi G_N(2p-\rho) e^{2\lambda}\\&&\nonumber+
2\left[\psi''+\psi'^2-\psi'\lambda'+\frac{2(\psi'-\lambda')}{r}+\frac{1-e^{2\lambda}}{r^2}\right].
\end{eqnarray}
Also, the trace equation gives us the following:
\begin{eqnarray}\label{phi}
&&\phi'^2+e^{2\lambda}=0.
\end{eqnarray}
From this differential equation we observe clearly that the auxiliary scalar field $\phi$ can be a complex function, i.e. $\phi\in\mathcal{C}$. We must not be worried about this solution, because even in this case, the reparametrization of the physical metric $g_{\mu\nu}$ does not yield a complex metric. 

Moreover, the hydrostatic (continuity) equation $\nabla_{\mu}T_{\nu}^{\mu}=0$ for $\nu=r$ yields 
\begin{eqnarray}
&&\frac{dp}{dr}=-(p+\rho)\psi'\label{p}.
\end{eqnarray}
This equation does not deviate from the one derived from GR because the auxiliary scalar field introduces no extra ghost degree of freedom. The only contribution comes from the metric and the fluid.
\par
In order to proceed with TOV equations, we redefine the metric function $\lambda$ in terms of a mass function $M(r)$ as follows:
\begin{eqnarray}
e^{-2\lambda}=1-\frac{2G_NM}{r}\label{dM}.
\end{eqnarray}
From this equation we can calculate the differential change in mass $dM$, as the mass which is stored in a layer with thickness $dr$ in the spacetime, as the following:
\begin{equation}\label{dM2}
\frac{G_N dM}{dr}=\frac{1}{2}\Big[1-e^{-2\lambda}(1-2r\lambda')\Big],
\end{equation}
with $M$ being the mass of the compact object. We can integrate $dM$ to find the total (active) mass inside a region $\mathcal{D}=\{r\in[0,R]\}$, with $R$ being the radius of the object. It was assumed that the mass enclosed inside the compact object is finite and a monotonic function of radial coordinate $r$.

The plan now is to rewrite Eqs.(\ref{tt})-(\ref{p}) in terms of $\{dp/dr,dM/dr,\rho,p\}$ in a dimensionless form. To do so, we make the following sequence of definitions: $M\to m M_{\odot}$, $r\to r_{g} r$, $\rho\to \rho M_{\odot}/r_{g}^3$, $p\to p M_{\odot}/r_{g}^3$ and $R\to R/r_g^2$, with $r_{g}=G_NM_{\odot}=1.47473km$ and $M_\odot$ is the mass of the Sun.

Such definitions yield (\ref{dM})-(\ref{dM2}) to convert to the following:
\begin{eqnarray}
\frac{d\lambda}{dr}=\frac{m}{r^2}\frac{1-\frac{r}{m}\frac{dm}{dr}}{\frac{2m}{r}-1}\label{dm}.
\end{eqnarray}
Using (\ref{p},\ref{dm}) in (\ref{tt}) yields, for the ($tt$) component:
\begin{eqnarray}
&&\label{eq11}
\frac{2m}{r^3}\frac{1-\frac{r}{m}\frac{dm}{dr}}{\frac{2m}{r}-1}-\frac{1}{r^2(1-\frac{r}{2m})}=\frac{8\pi \rho}{1-\frac{2m}{r}},
\end{eqnarray}
while the ($rr$) and trace equations are, respectively:
\begin{eqnarray}
&& \label{eq22}
\left(\frac{p'}{p+\rho}\right)^2-\left(\frac{p'}{p+\rho}\right)'+\frac{m}{r^2}\frac{1-\frac{r}{m}\frac{dm}{dr}}{\frac{2m}{r}-1}\frac{p'}{p+\rho}\\&&\nonumber-\frac{p'}{r(p+\rho)}-\frac{2m}{r^3}\frac{1-\frac{r}{m}\frac{dm}{dr}}{\frac{2m}{r}-1}+\frac{1}{2r^2(1-\frac{r}{2m})}=\frac{8\pi (2p-\rho)}{1-\frac{2m}{r}},
\end{eqnarray}
\begin{eqnarray}
&&  
\frac{d\phi}{dr}=\frac{r_g}{\sqrt{\frac{2m}{r}-1}}
\label{eq33}.
\end{eqnarray}

We can combine Equations (\ref{p},\ref{eq11},\ref{eq22},\ref{eq33}) and obtain:
\begin{eqnarray}
&&\phi'' = -\frac{r_g^2}{\phi'}\frac{4 r^3\pi\rho-m}{4m^2-4 m r+ r^2}\label{dyn1},\\&&
\psi''\big(-r^3+2r^2m) +m(\psi' r+2\psi'^2 r^2-1)-\psi' r^2\\&&\nonumber-\psi'^2 r^3-4\psi' r^4\pi\rho+16\pi r^3 p=0\label{dyn2},\\&& m' = 4\pi\rho r^2\label{dyn3},\\&& p' = -\psi'(p+\rho)\label{dyn4},
\end{eqnarray}
with the following constraints: $2\phi'^2m-r\phi'^2-r_g^2r = 0,r\neq 2m, p+\rho\neq 0, \phi'\neq0$.

An alternative elimination process is to eliminate $m$ first, then $p,\rho,\phi$ and finally $\psi$. From this procedure we obtain the following system of differential equations: 
\begin{eqnarray}
&&m =\frac{r}{2}\frac{\phi'^2+r_g^2 }{\phi'^2},\\&&
-32p\pi r^2\phi'^3-2\phi' r_g^2\psi'^2r^2
+\phi' r_g^2+\phi'^3-2\phi' r_g^2r^2\psi'
\\&&\nonumber-2\phi'r_g^2\psi'r+2\psi'
r^2\phi''r_g^2=0,\\&&
\rho = \frac{1}{8}\frac{\phi'^3+\phi'r_g^2-2r\phi''r_g^2}{\pi r^2\phi'^3},
\\&&
2\phi''\phi' r^3\psi'r_g^2  -6\psi' r^3\phi''^2r_g^2+5\phi'^4
r \psi'+7\phi'^2 r\psi'r_g^2\\&&\nonumber-4\phi'r^2\psi'\phi'r_g^2
+6\phi' r^3\phi''
r_g^2\psi'^2-6\phi'^2r^3r_g^2
\psi'\psi''\\&&\nonumber-2\phi' rr_g^2\phi''
-2\phi'^2 r^2r_g^2\psi''-2\phi'^2
r^3r_g^2\psi''\\&&\nonumber-2\phi'^2\psi'^3
r^3r_g^2-2\phi'^2r_g^2\\&&\nonumber-2\phi'^4
+6\phi'r_g^2r^3\psi''\phi''-2\phi'^2
r^2r_g^2\psi'^2=0,
\end{eqnarray}
with the following constraints 
\begin{eqnarray}
&&
-5\phi'^3-5\phi'r_g^2+8r\phi''r_g^2
+2\phi'r_g^2\psi'^2r^2\\&&\nonumber+2\phi'r_g^2r^2
\psi''+2\phi'r_g^2\psi' r-2\psi'r^2\phi''r_g^2 \neq0 ,\\&& \psi' \neq 0,\phi'\neq 0.
\end{eqnarray}
But let us now revert to (\ref{dyn1}-\ref{dyn4}), which gives us a system of first order autonomous differential equations (see \cite{Leon:2014rra} for a review of dynamical systems approach in cosmology):
\begin{eqnarray}
&&\phi'=\zeta,\\&&
\zeta' = -\frac{r_g^2}{\zeta}\frac{4 r^3\pi\rho-m}{4m^2-4 m r+ r^2}\label{dyn11},\\&&
\psi'=\mu,\\&&
\mu'(-r^3+2r^2m) +m(\mu r+2\mu^2 r^2-1)-\mu r^2\\&&\nonumber-\mu^2 r^3-4\mu r^4\pi\rho+16\pi r^3 p=0\label{dyn22},\\&& m' = 4\pi\rho r^2\label{dyn33},\\&& p' = -\mu(p+\rho)\label{dyn44}.
\end{eqnarray}
These equations can be written in the form of an autonomous set of first order differential equations:
\begin{eqnarray}
&&\vec{X}'=f(\vec{X};r),\vec{X}=(\phi,\zeta,\psi,\mu,m,p).
\end{eqnarray}
We use (\ref{dyn11}-\ref{dyn44}) to find all stationary points of the system. Such points are the solutions of $f(\vec{X_c},r_0)=0$, with $r_0$ being an arbitrary value of $r$. For our problem we can solve this and find the following stationary point:
\begin{eqnarray}
&&\vec{X_c}=(\phi_c,0,\psi_c,0,0,0).
\end{eqnarray}
We can now linearize the system of equations (\ref{dyn11}-\ref{dyn44}) in the vicinity of this critical point. The technique is to make $X=X_c+\delta \vec{X}$ and expand the function $f$ up to the first order of $\delta \vec{X}$. In this case, thanks to the stationary condition, we will obtain the following set of first order ``linear" differential equations for the perturbations:
\begin{eqnarray}
&&\frac{d}{dr}\delta \vec{X}=J \delta\vec{X}.
\end{eqnarray}
Here, the matrix of the coefficients is called ``the Jacobian matrix" and is defined by $J=\frac{Df}{DX}|_{X_c}$. For our case we calculate it explicitly as follows:
\begin{eqnarray}
J= \left[ \begin {array}{cccccc} 0&1&0&0&0&0\\ \noalign{\medskip}0&0&0&0
&{\frac {{r_{{g}}}^{2}}{X_{{c}}{r}^{2}}}&0\\ \noalign{\medskip}0&0&0&1
&0&0\\ \noalign{\medskip}0&0&0&-{r}^{-1}&-{r}^{-3}&16\,\pi 
\\ \noalign{\medskip}0&0&0&0&0&4\,\pi \,\rho_{{p}}{r}^{2}
\\ \noalign{\medskip}0&0&0&-\rho \left( 0 \right) &0&0\end {array}
\right],
\end{eqnarray}
with $X_c\to0$ and by $\rho_p$ we mean $d\rho/dp$. 

To classify the stationary point as stable or unstable, we need to compute all eigenvalues of this  real valued matrix. An eigenvalue of $J$ is a scalar, such that there exists a vector (the corresponding eigenvector) for which $J\vec{V}=\lambda \vec{V}$. The characteristic polynomial is given by the following six order algebraic equation:
\begin{eqnarray}
&&{\lambda}^{3} [{\lambda}^{3}r+{\lambda}^{2}-16\,\rho \left( 0
\right) \pi \,\lambda\,r-4\,\rho \left( 0 \right) \pi \,\rho_{{p}}
]
=0.
\end{eqnarray}
In the above equation we suppose $\rho=\rho(p)$ so that $\rho(0)$ means $\rho(p=0)$. The solutions degenerate for $\lambda_{i=1,2,3}=0$, so the system is unstable under infinitesimal perturbations $\vec{X}\to \vec{X}+\delta\vec{X}$. Another real valued root is given by the following:
\begin{eqnarray}
&&\lambda_R=\frac{1}{3}\,{\frac {{C}^{4}+48\,\rho \left( 0 \right) \pi \,{r}^{2}+1
		-{C}^{2}}{{C}^{2}r}},
\end{eqnarray}
with $C$ defined as
\begin{eqnarray}
&& C^6=72\,\rho \left( 0 \right) \pi \,{r}^{2}54\,\rho \left( 0
\right) \pi \,\rho_{{p}}{r}^{2}-1+6\,Br,
\end{eqnarray}
and
\begin{eqnarray}
&&\frac{B^2}{3\,\rho \left( 0 \right) \pi}= -1024\, \rho(0)^{2}{\pi }^{2}{r}^{4}-16\,\rho \left( 0
\right) \pi \,{r}^{2}\\&&\nonumber+72\,\rho \left( 0 \right) \pi \,\rho_{{p}}{r}^{
	2}+27\,\rho \left( 0 \right) \pi \,{\rho_{{p}}}^{2}{r}^{2}-\rho_{{p}}.
\end{eqnarray}

\section{Solutions for different star models}\label{sec:ns}
We will solve Eqs.(\ref{p},\ref{eq11},\ref{eq22},\ref{eq33}) numerically with an auxiliary EoS $p=p(\rho)$ for each of the compact star models below. Before starting to solve such equations for QS models, it is worth stressing that QSs are, indeed, likely to exist. In \cite{moraes/2014b}, for instance, it is presented a form of probing the existence of compact binaries containing at least one QS from the observation of gravitational waves. Once QSs existence is probed, the ground state of matter will be understood as that of strange matter, i.e., a quark fluid with about equal numbers of up, down and strange quarks.

\subsection{QS model with linear EoS}
Quark stars have been considered also in  other types of modified gravities, mainly in $f(R)$ gravity it has been well investigated for several models 
\cite{ Phys.Lett. B742 (2015) 160}. In Ref. \cite{ Phys.Lett. B742 (2015) 160}
quark star models with realistic equation of state investigated using a   nonperturbative method. One of important feature is that in this model, the mass-radius relation has been obtained numerically for a type of modified gravity with action  $f(R) = R + \alpha R^2$ which is considered as an alternative model to describe inflation. Actually in $f(R)$ gravity based on this analysis it has been determined the  value of central density, the unique value of central curvaturein an asymptotically Ricci flat geometry. An alternative description developed using scalar-tensor equivalence of $f(R)$ gravity where the vital fine-tuning on Ricci scalar $R$ is equivalent to the fine-tuning on the scalar field $\phi$. The role of coupling constant $\alpha$ has been studied and it has been proven that the gravitational mass of the star increases with increasing $\alpha$. 
\par
Inspiring from this refrence, in this section we will study quark star models within MG scenarion.
Let us start solving the system (\ref{p},\ref{eq11},\ref{eq22},\ref{eq33}) by adopting the following quark matter EoS \cite{ Phys.Lett. B742 (2015) 160},\cite{N. Stergioulas}:
\begin{eqnarray} 
p=c(\rho-4B),c = 0.28, 0.98 < B < 1.52. 
\end{eqnarray}
Solutions for such a system are given, analytically, by the following elementary expressions:

\begin{eqnarray}
&&\phi(r)=2r_gr+C_1,\\
&&m(r)=\frac{5}{8}r,\\
&&\rho(r)=\frac{5}{96\pi }\frac{1}{r^2},\\
&&p(r)=\frac{5c}{96\pi }\Big(\frac{1}{r^2}-\frac{1}{R^2}\Big),\\
&&\psi(r)=\frac{c}{c+1}\ln\Big[\frac{r^2}{(1+c)R^2-cr^2}\Big],\ \ \frac{r}{R}<2.139.
\end{eqnarray}
It is noted, for the density function $\rho(r)$ in the vicinity of $r=0$, a singularity for such a QS model. Furthermore, the metric function $\psi(R)=0$.

\subsection{QS model with exponential EoS}
An empirical EoS is proposed in \cite{momeni/2015} as the following:

\begin{eqnarray}
\frac{p}{p_c}= a (1- e^{-b\frac{\rho}{\rho_c}}).\label{eos}
\end{eqnarray}
Here, $p_c=\rho_c=r_g^3/M_{\odot}$, and $a,b$ are constants. Low density profiles increase significantly the pressure and makes the \emph{hardenability}\footnote{The depth up to which a star is hardened during a matter collapse process.} of the star. The QS with linear EoS $p=A\rho+B$ can be considered a special case of (\ref{eos}) when $\rho\ll \rho_c$. 

\begin{table}[h]
	\caption{The parameters of EoS for three different stars: $\frac{p}{p_c}= a (1- e^{-b\frac{\rho}{\rho_c}})$. }
	\begin{center}
		\begin{tabular}{|c|c|c|c|c|}
			\hline
			Strange star candidates          & a       & $\sigma_a$ & b        & $\sigma_b$ \\ \hline
			Her X-1           & 0.45526 & 0.0198     & 0.0152   & 0.00124    \\ \hline
			SAX J 1808.4-3658 & 1.2242  & 0.0563     & 0.0191   & 0.00162    \\ \hline
			4U 1820-30        & 1.1255  & 0.0408     & 0.0243   & 0.00163    \\ \hline
		\end{tabular}
	\end{center}
\end{table}

We list in Table 1 above the parameters for three astrophysical objects. In the 4U 1820-30 object, the effects of high pressure are much more important than the same effects on the other objects. Also we tested the parameters of EoS (\ref{eos}) for several initial conditions. We did not observe any significant difference among the models. The schematic form of EoS (\ref{eos}) remains the same. In fact, it can be considered as a generalized exponential virial EoS proposed in the literature \cite{Kenneth}:

\begin{eqnarray}\label{Kenneth}
&&\frac{p}{\rho k_B T}=\exp\Big(\Sigma_{m=2}^{\infty}K_m \rho^{m-1}\Big).
\end{eqnarray}
In the EoS above, the coefficients $K_m$ and $B_n$ can be computed in terms of the Virial coefficients. Also, $k_B$ is the Boltzmann constant and $T$ is the temperature. However, this similarity is just formal. There is a significant 
difference between our EoS (\ref{eos}) and exponential Virial EoS (\ref{Kenneth}). It is related to the  pressure of the background $p_b\sim p_c$. In Kenneth's EoS (\ref{Kenneth}), because it works mainly for gases with
exponential virial EoS, the pressure of the background $p_b=0$. Meanwhile in our compact object model, we have a non zero $p_b$
value. Another difference is that we constructed EoS (\ref{eos}) numerically from the gravitational
field equations, whereas exponential virial EoS (\ref{Kenneth}) was made from
high-density fluid. 

We solve (\ref{p},\ref{eq11},\ref{eq22},\ref{eq33}) numerically. The results are shown for Her X-1,4U 1820-30 and SAX J 1808.4-3658 in the following.

\subsubsection{Her X-1}
For Her X-1, $m\sim 0.88$, $R\sim 7.7$ and we expect the pressure to vanish at $r\sim R$ \cite{rahman1,rahman2}. Our numerical simulations for (\ref{p},\ref{eq11},\ref{eq22},\ref{eq33}) are shown in Figs.1 and 2. From Fig.1, we see that the mass function  always increases. Scalar field $\phi$ and metric function $\psi$ also are monotonic-increasing functions of radial coordinate $r$. But the situation is different for pressure $p$ and energy $\rho$, which are monotonic-decreasing. At large distances $r\sim 8$, the pressure $p$ vanishes.

\begin{figure}[h!]
	\centering
	\includegraphics[width=0.7\textwidth]{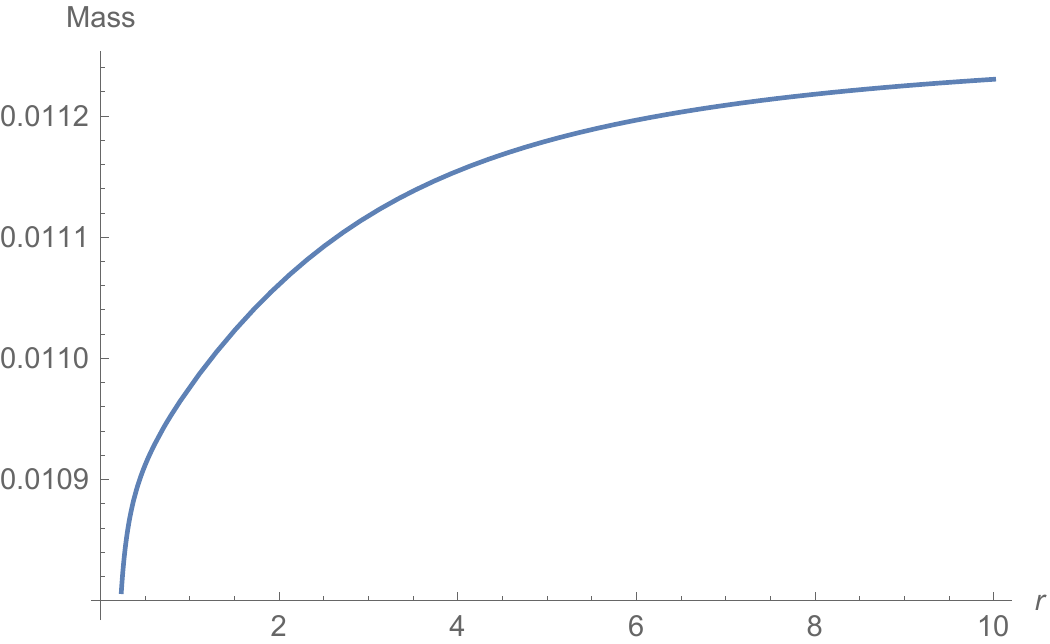}\caption{The graphical plot of mass vs. radius for Her X-1 with EoS (\ref{eos}).}\label{herx-mass}
\end{figure}

\begin{figure}[h!]
	\centering 
	\includegraphics[width=.7\textwidth,clip]{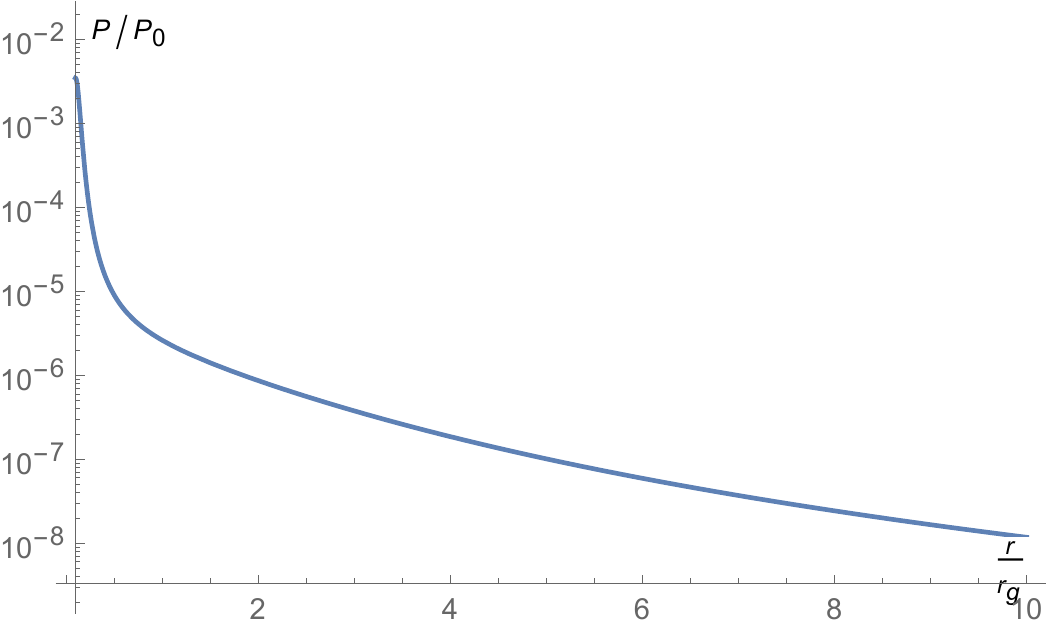}
	\hfill
	\includegraphics[width=.7\textwidth,origin=c]{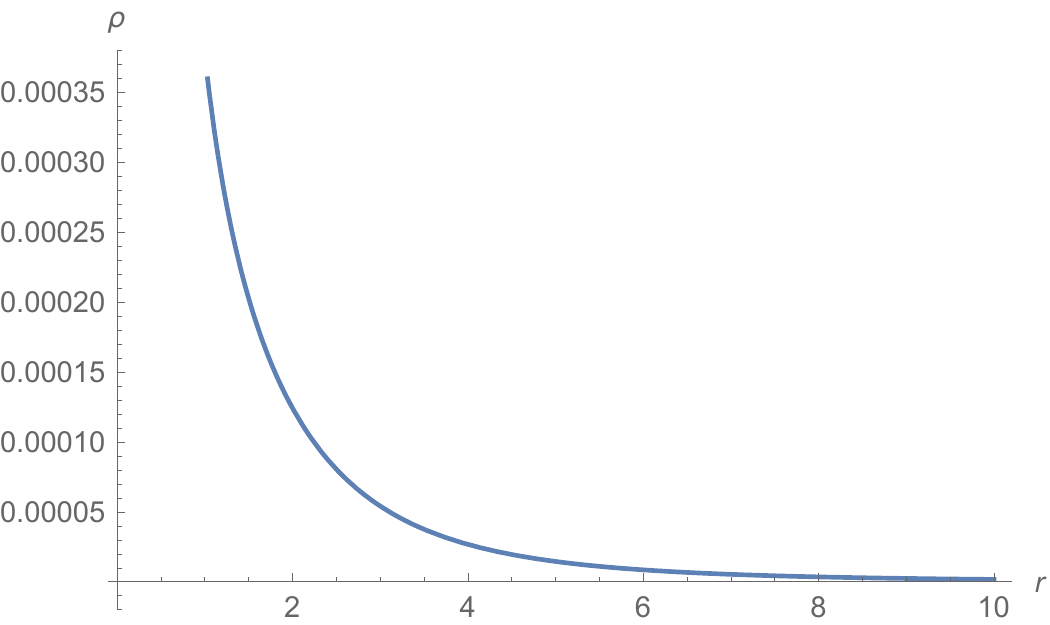}
	\hfill
	\includegraphics[width=.7\textwidth,clip]{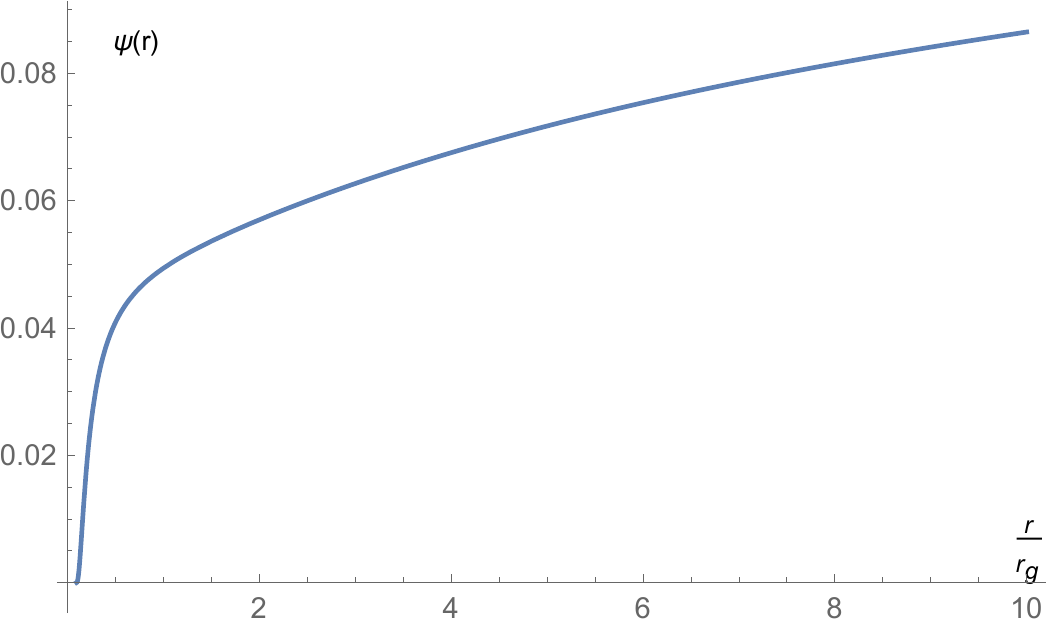}
	\hfill
	\includegraphics[width=.7\textwidth,origin=c]{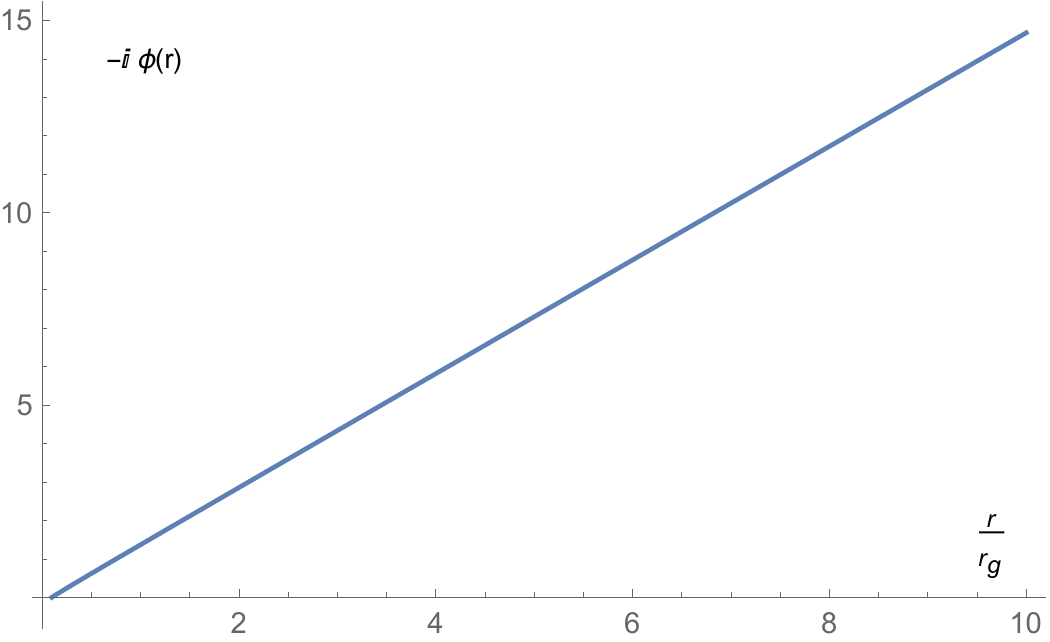}
	\caption{The graphical normalized plots for pressure $p$, density profile $\rho$, metric function $\psi$ and scalar field $\phi$ vs. radius for Her X-1 with EoS (\ref{eos}).}\label{p,rho,psi,phi,HerX}
\end{figure}

\subsubsection{4U 1820-30}

For 4U 1820-30, $m\sim 2.25$, $R\sim 10$ and we expect that the pressure vanishes for $r\sim R$ \cite{rahman3}. Our numerical simulations for (\ref{p},\ref{eq11},\ref{eq22},\ref{eq33}) are shown in Figs.3 and 4. From Fig.3, we observe that the mass function  always increases. Scalar field $\phi$ and metric function $\psi$ are also monotonic-increasing functions of the radial coordinate $r$. But the situation is different for pressure $p$ and energy $\rho$. Both of these functions are monotonic-decreasing. Moreover, at large distances $r\sim 10$, the pressure $p$ vanishes. 

\begin{figure}[tbp]
	\centering
	\includegraphics[width=0.7\textwidth]{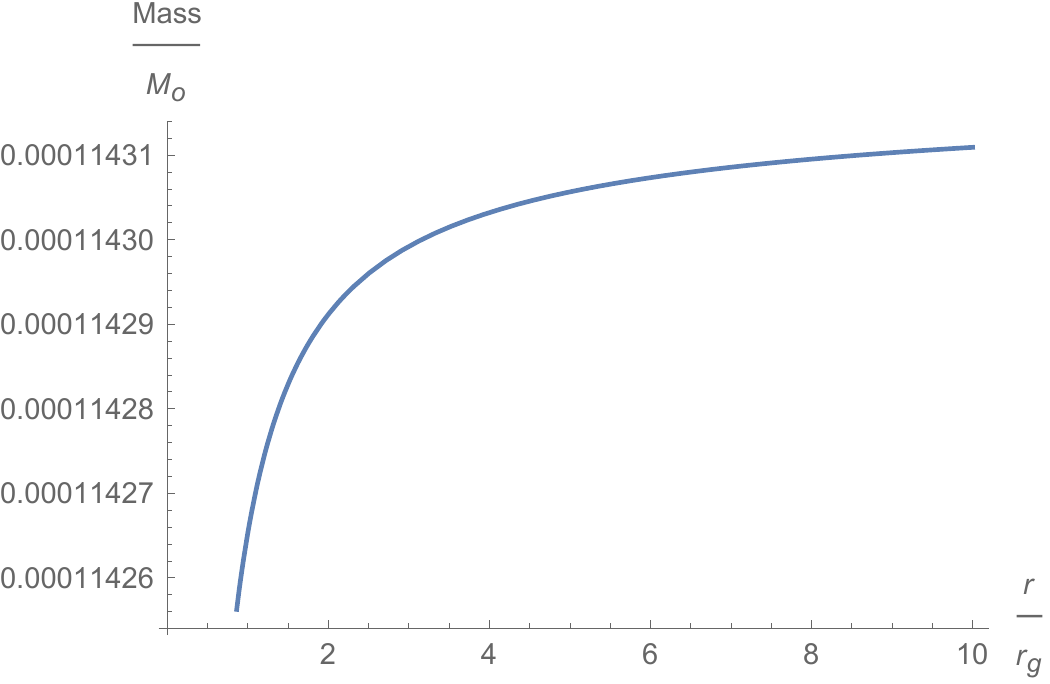}\caption{The graphical plot of mass vs. radius for 4U 1820-30 with EoS (\ref{eos}). We use the following initial conditions: $m(0.01)=0.0001$, $\rho(0.01)=10$, and $\rho'(0.01) =0$.}\label{4U 1820-30-mass}
\end{figure}

\begin{figure}[tbp]
	\centering 
	\includegraphics[width=.7\textwidth,clip]{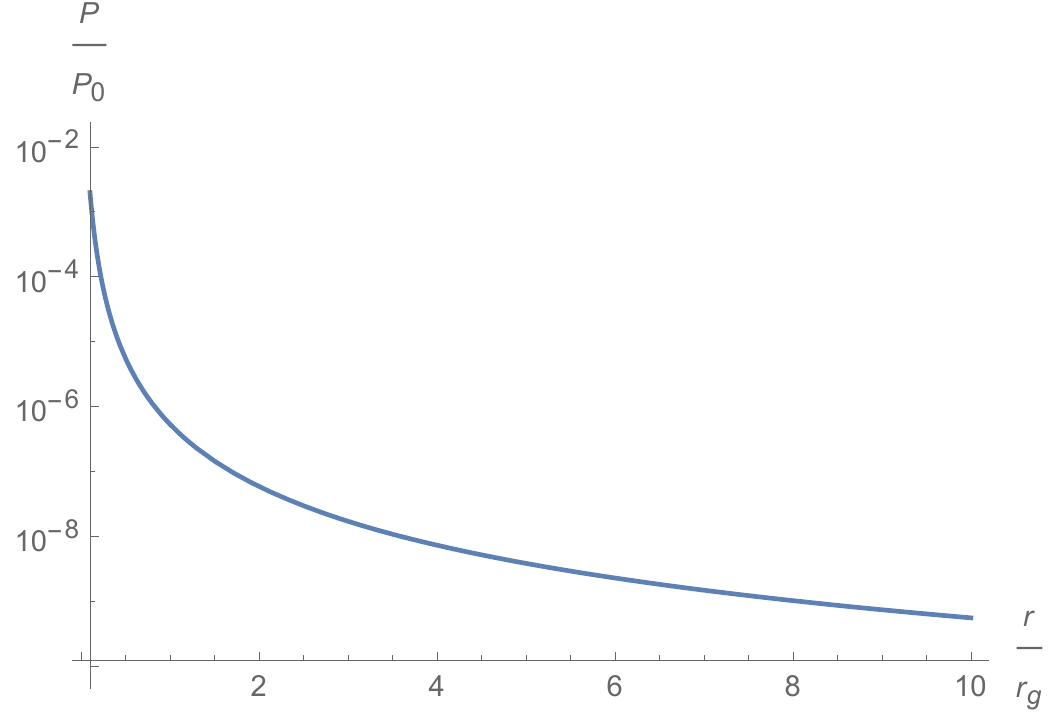}
	\hfill
	\includegraphics[width=.7\textwidth,origin=c]{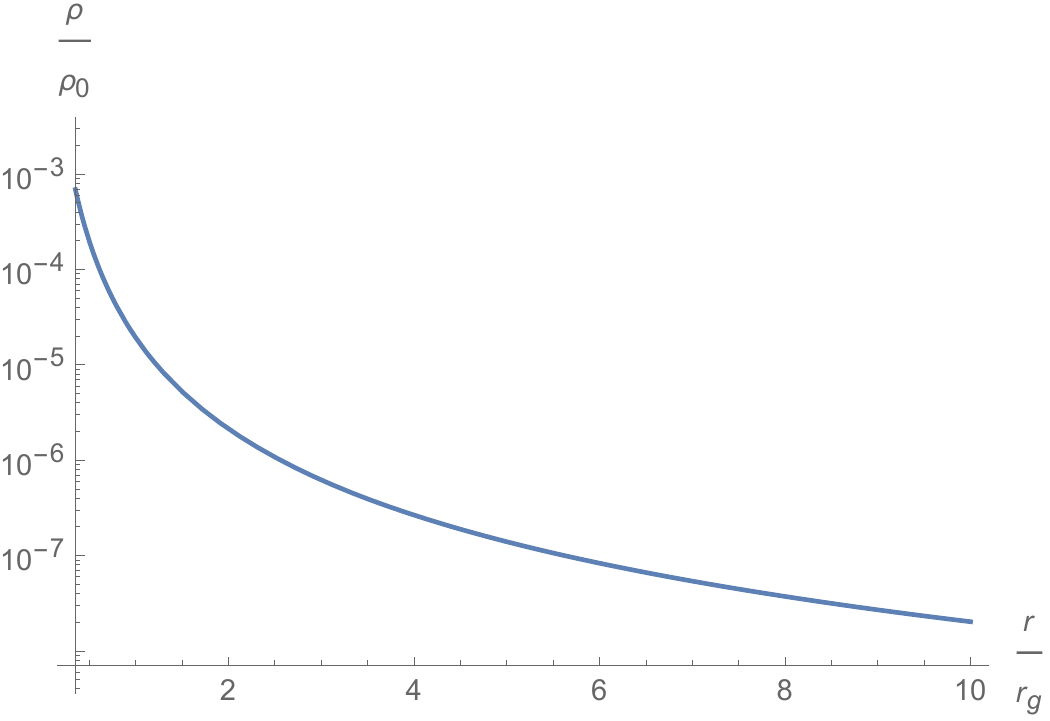}
	\hfill
	\includegraphics[width=.7\textwidth,clip]{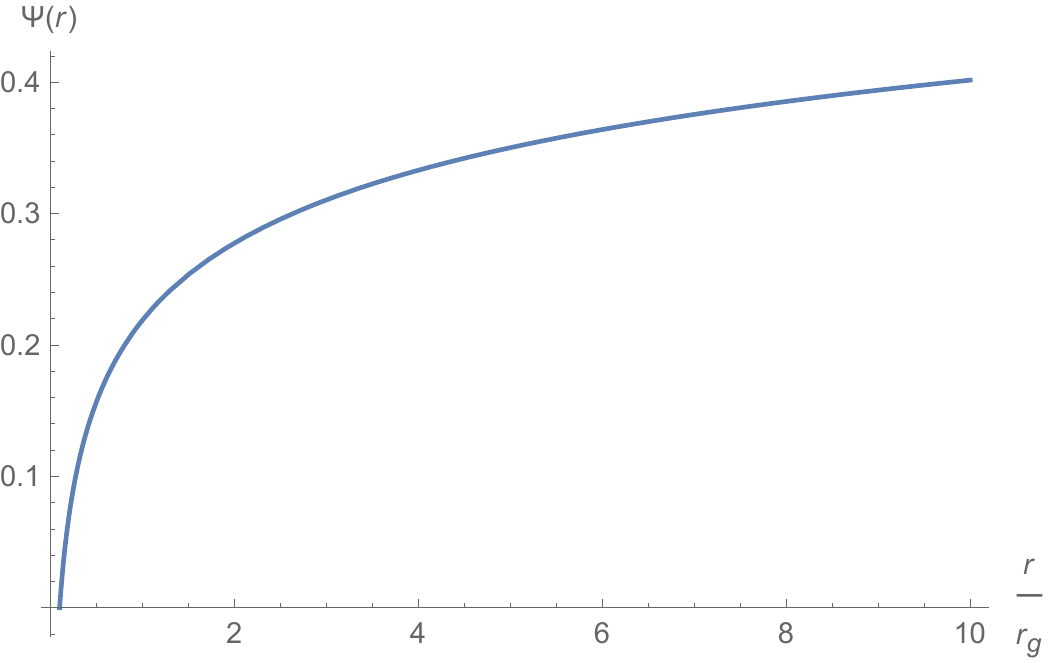}
	\hfill
	\includegraphics[width=.7\textwidth,origin=c]{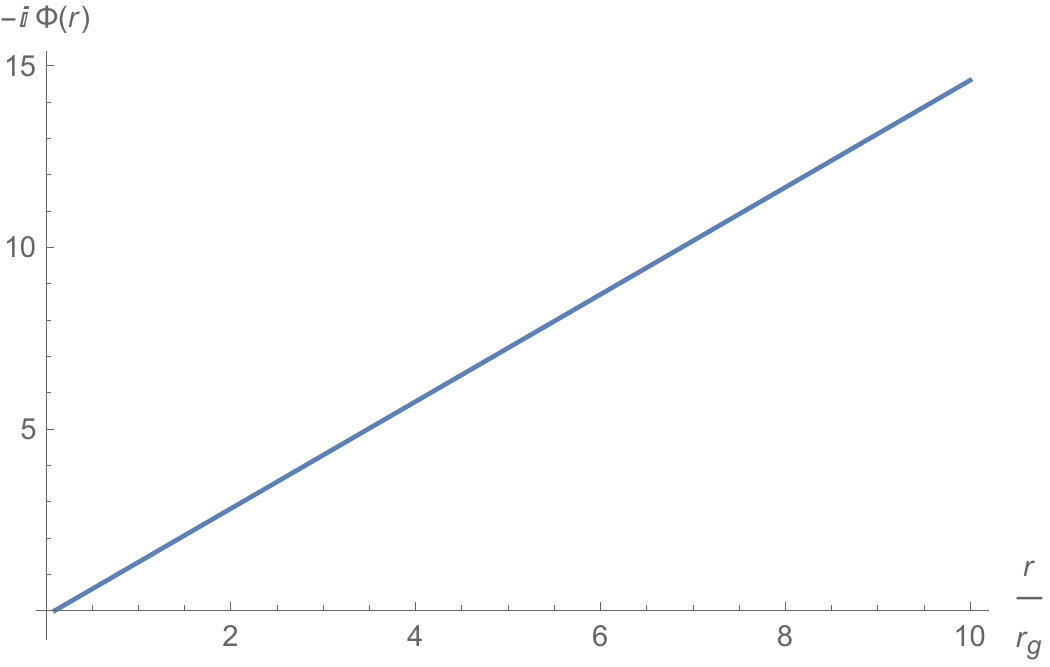}
	\caption{The graphical normalized plots for pressure $p$, density profile $\rho$, metric function $\psi$ and scalar field $\phi$ vs. radius for 4U 1820-30 with EoS (\ref{eos}).}\label{p,rho,psi,phi,4U 1820-30}
\end{figure}

\subsubsection{SAX J 1808.4-3658}
For SAX J 1808.4-3658, $m\sim 1.435$, $R\sim 7.07$ and we expect that the pressure vanishes for $r\sim R$ \cite{rahman3}. Our numerical simulations for (\ref{p},\ref{eq11},\ref{eq22},\ref{eq33}) are shown in Figs.5 and 6. From Fig.5, we observe that the mass function  always increases. Scalar field $\phi$ and metric function $\psi$ are also monotonic-increasing functions of radial coordinate $r$, as for the other objects which are monotonic-decreasing. At large distances $r\sim 10$, the pressure $p$ vanishes.

\begin{figure}[tbp]
	\centering
	\includegraphics[width=0.7\textwidth]{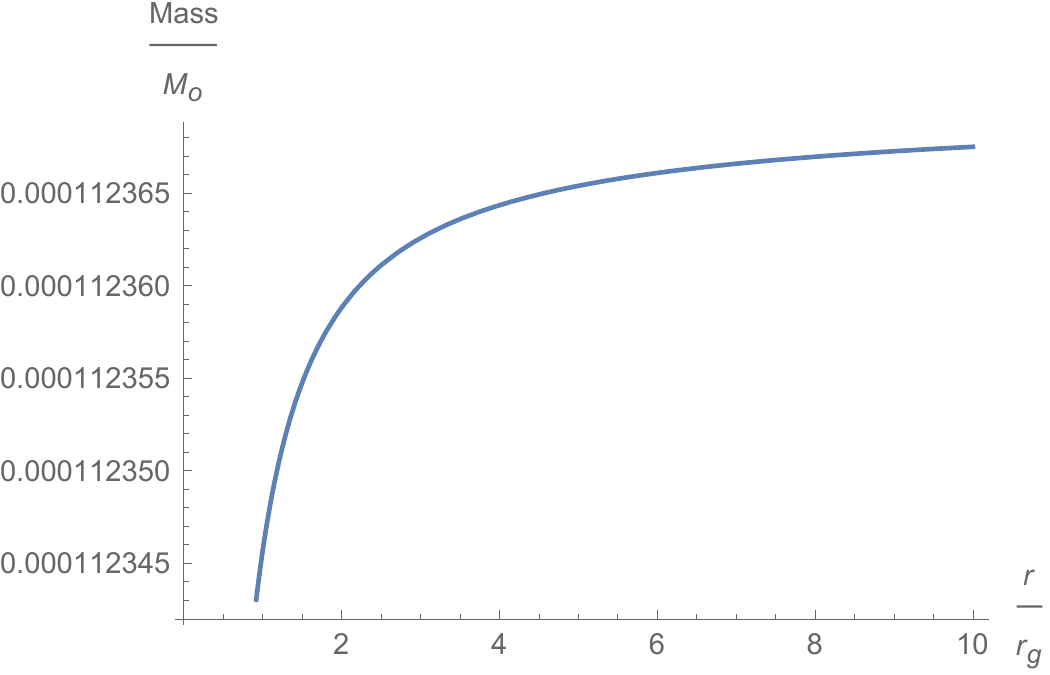}\caption{The graphical plot of mass vs. radius for SAX J 1808.4-3658 with EoS (\ref{eos}).}\label{SAX J 1808.4-3658-mass}
\end{figure}

\begin{figure}[tbp]
	\centering 
	\includegraphics[width=.7\textwidth,clip]{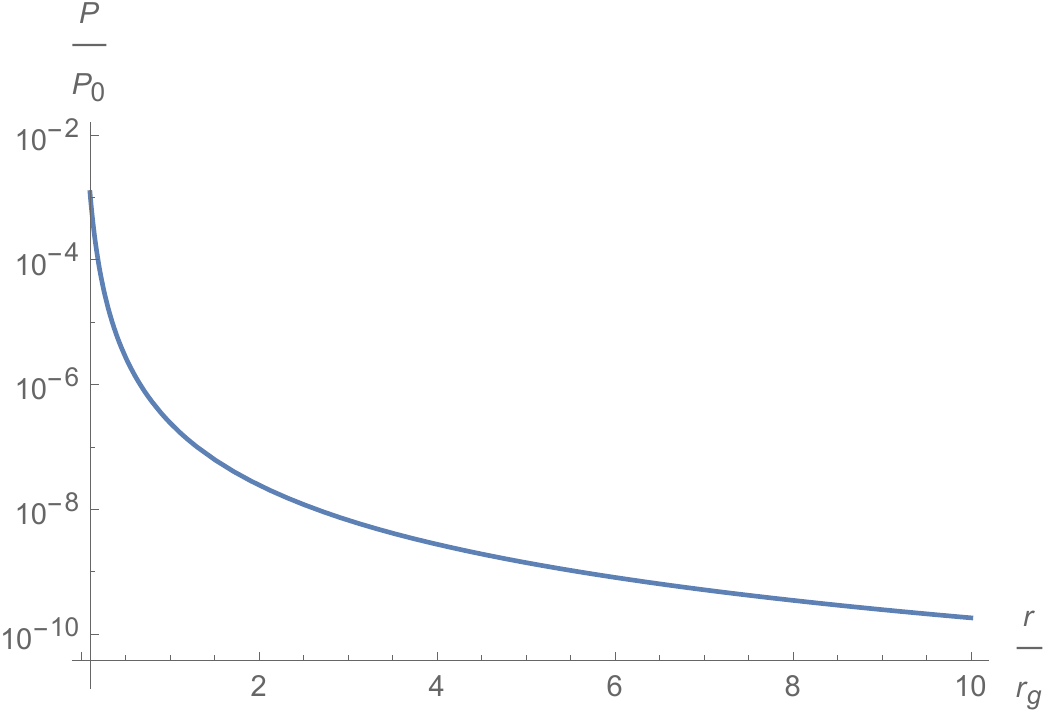}
	\hfill
	\includegraphics[width=.7\textwidth,origin=c]{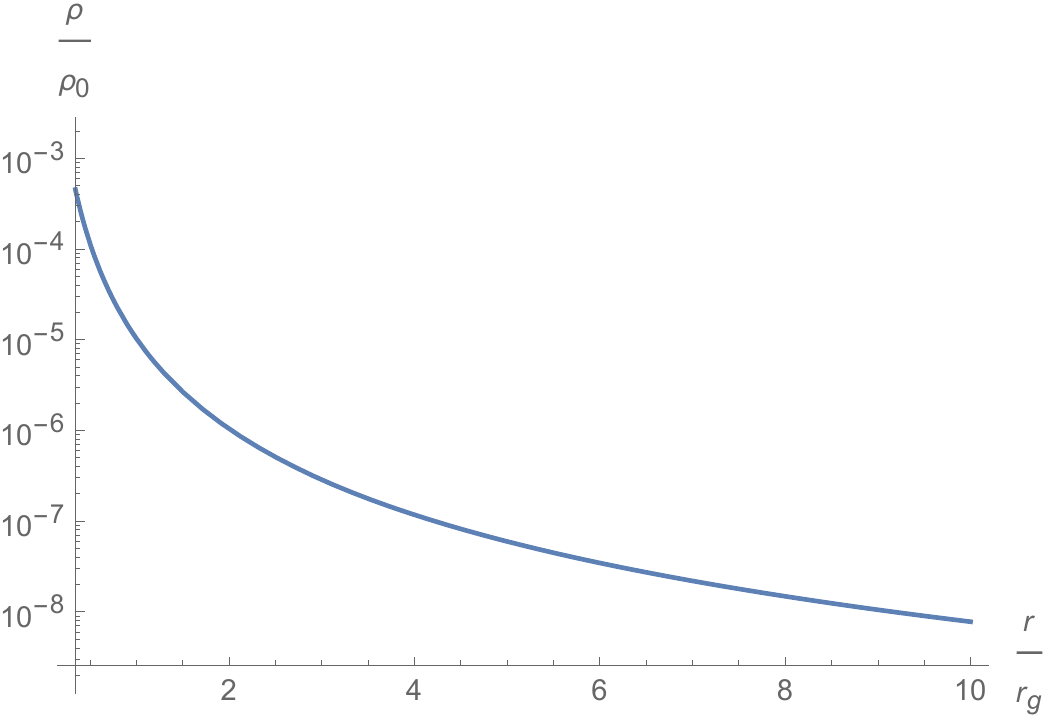}
	\hfill
	\includegraphics[width=.7\textwidth,clip]{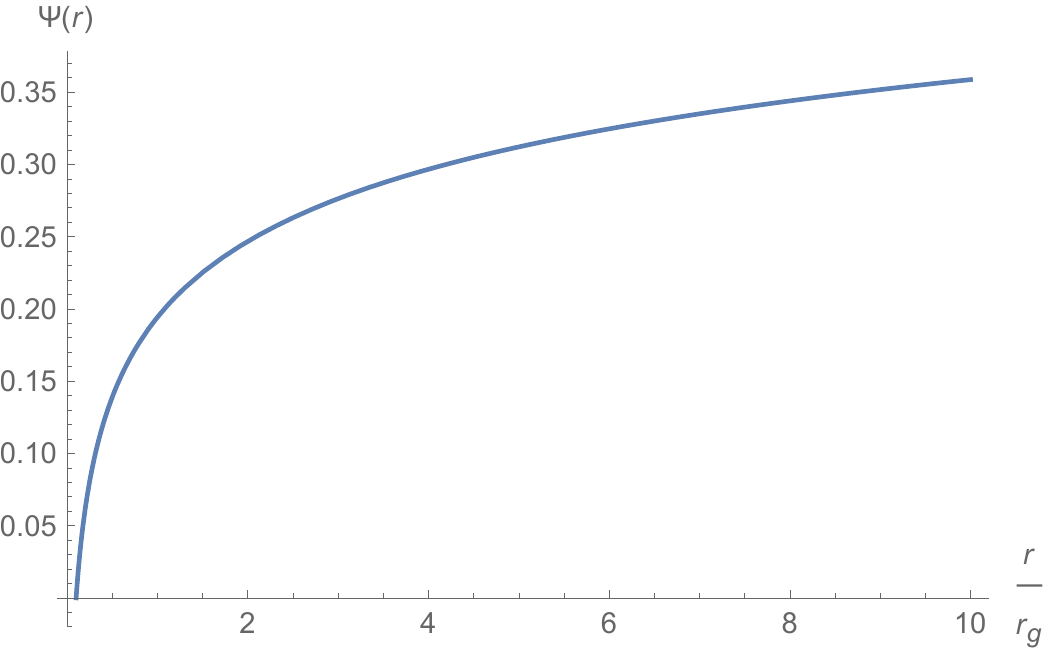}
	\hfill
	\includegraphics[width=.7\textwidth,origin=c]{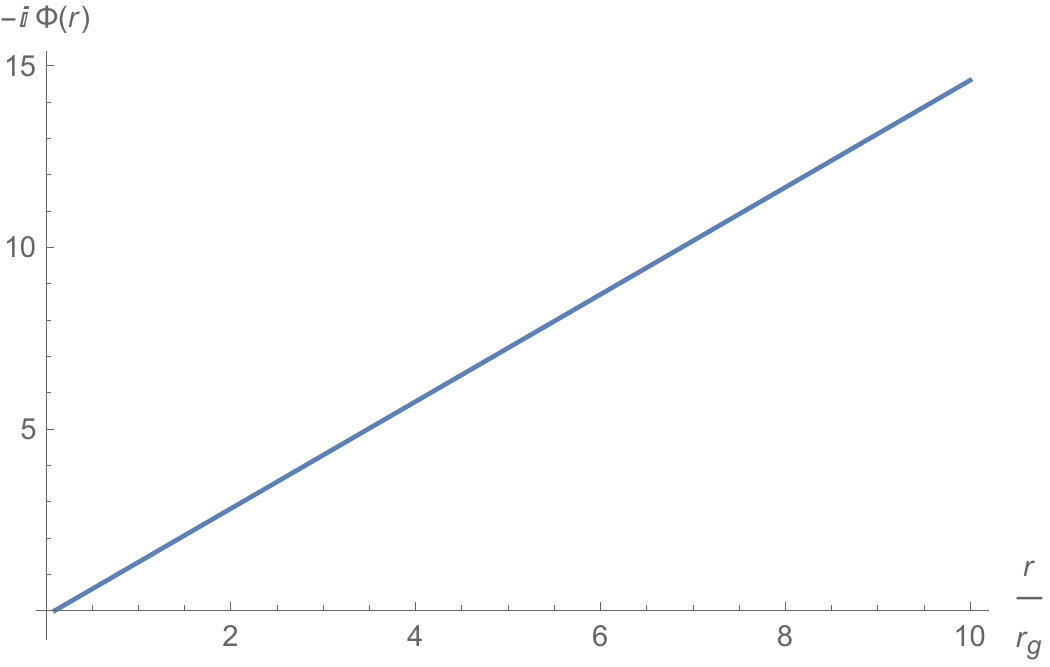}
	\caption{The graphical normalized plots of pressure $p$, density profile $\rho$, metric function $\psi$ and scalar field $\phi$ vs. radius for SAX J 1808.4-3658 with EoS (\ref{eos}).}\label{p,rho,psi,phi,SAX J 1808.4-3658}
\end{figure}

\subsection{NS with polytrope EoS}
A polytrope EoS for NS was proposed in the following form \cite{Lattimer:2000nx}:
\begin{eqnarray}
&&p = K\rho^{1+\frac{1}{\Gamma}},
\end{eqnarray}
with $K$ and $\Gamma$ being constants.

If we integrate the first law of thermodynamics, we obtain:
\begin{eqnarray}
&&\rho=p+\Big(\frac{p}{K}\Big)^{\frac{\Gamma}{\Gamma+1}}.
\end{eqnarray}
By setting $K=1$ and $\Gamma=2$, this EoS leads to compact objects with accepted mass and radius for NSs
\cite{Lattimer:2000nx}. In fact, it has been recently used in Horndeski's models of NSs \cite{Cisterna:2015yla}.
Our numerical simulations for (\ref{p},\ref{eq11},\ref{eq22},\ref{eq33}) are shown in Figs.7 and 8. From Fig.7, we observe that the mass function increases at some points and remain constant at others. Therefore mass with polytropic EoS is not a monotonic increasing function. Scalar field $\phi$ and metric function $\psi$ are not monotonic-increasing functions of radial coordinate $r$. For scalar field, the function is linearly dependent on $r$. The metric function $\psi$ is increasing for $r\leq 2.5$, monotonic-decreasing for $2.5<r<5.5$ and it reaches a local minimum at $r\sim 5.5$. It is monotonic-increasing for $r\sim (5.5 -9.5)$ and at the radius $r\sim9.5$ it becomes discontinuous. In Fig.8 we draw the graphics for density and pressure. The density $\rho$ is not a monotonic increasing function. It has two minima at $r\sim 5,11.75$ and vanishes at distances larger than $r\geq 17$. This would, of course, create a periodic behavior, which would be determinable for the pressure and density of any particular point (say the center) of the star, but the pressure $p$ would be smaller for a point near the star radius and greater for a point farther from the star. In fact, for $r\sim 8$ the pressure $p$ vanishes.  

\begin{figure}[tbp]
	\centering
	\includegraphics[width=0.7\textwidth]{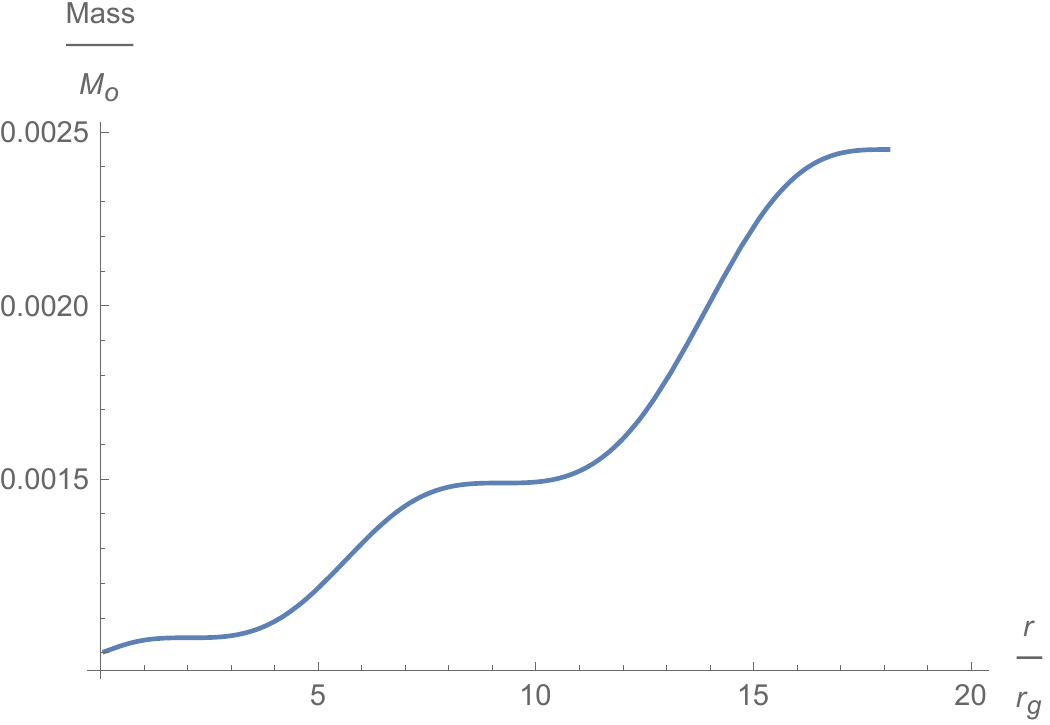}\caption{The graphical plot of mass vs. radius for the NS with polytropic EoS.}\label{Polytrope-mass}
\end{figure}

\begin{figure}[tbp]
	\centering 
	\includegraphics[width=.7\textwidth,clip]{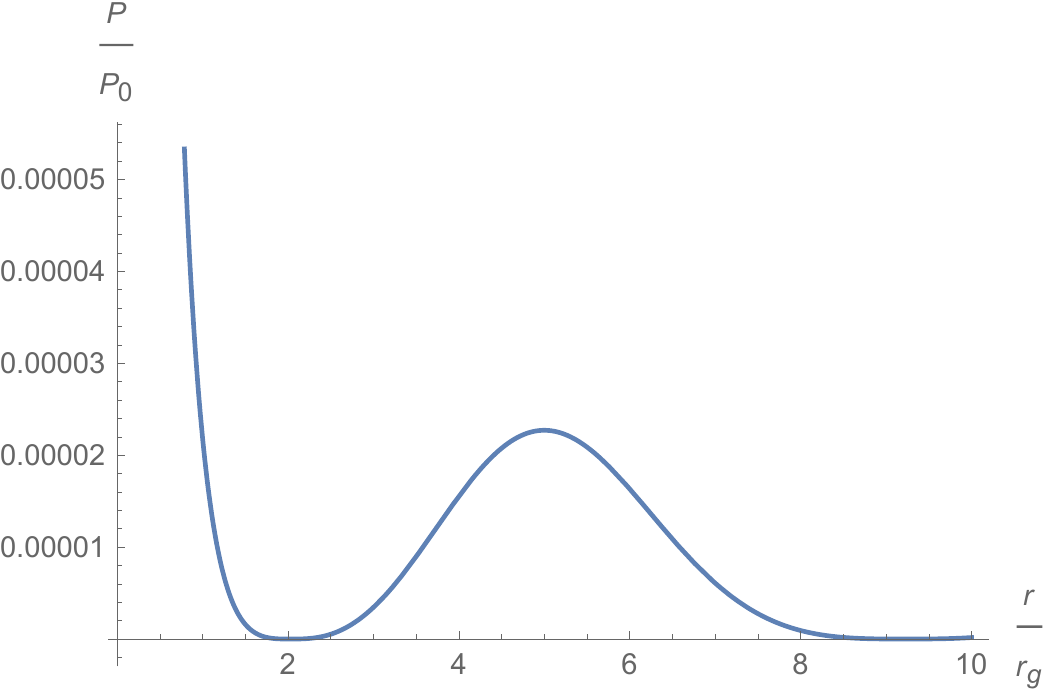}
	\hfill
	\includegraphics[width=.7\textwidth,origin=c]{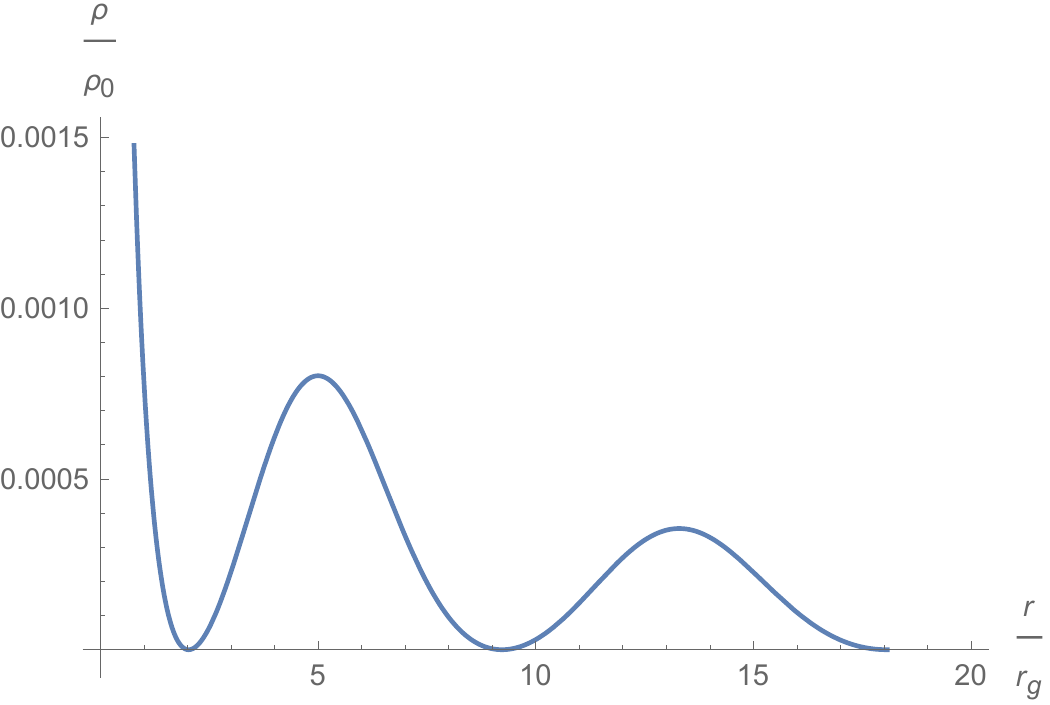}
	\hfill\includegraphics[width=.7\textwidth,clip]{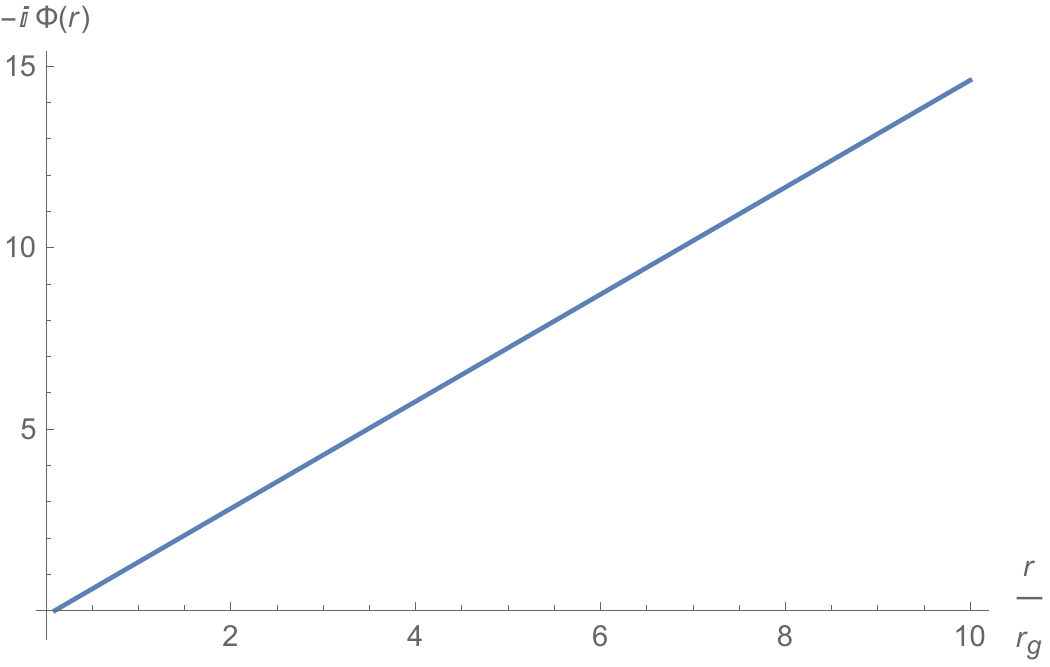}
	\hfill
	\includegraphics[width=.7\textwidth,origin=c]{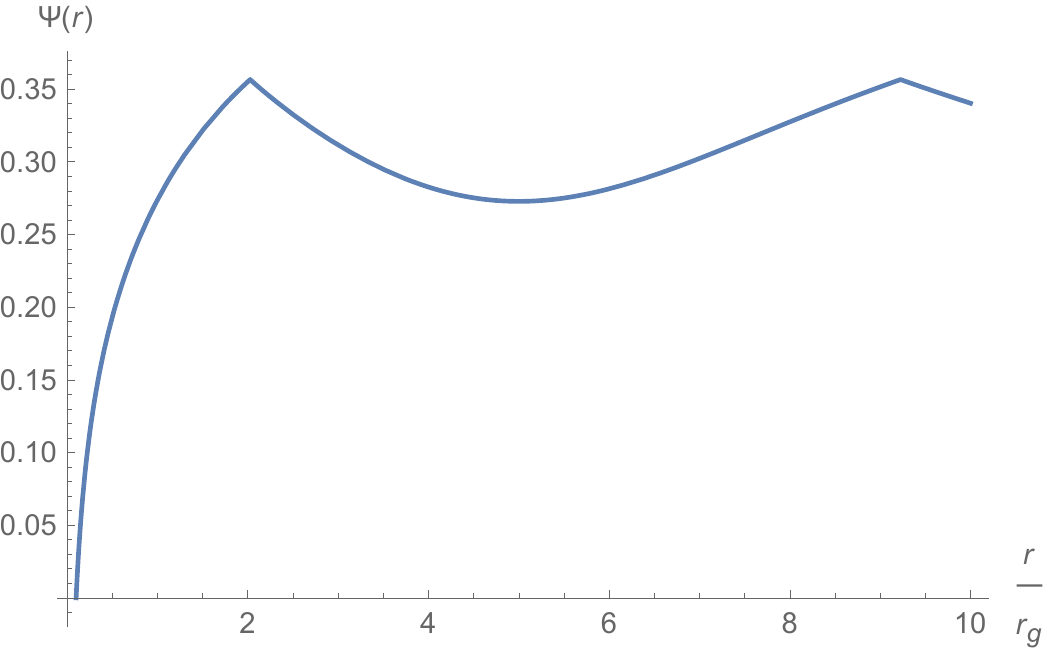}
	\caption{The graphical normalized plots of pressure $p$, density profile $\rho$, metric function $\psi$ and scalar field $\phi$ vs. radius for the NS with polytropic EoS. The curves for $p$ and $\rho$ are periodic, and cut the $r$-axis at the point $r\sim 2.5$.  The maximum value of the pressure is $p_{max}\sim 2.5\times 10^{-5}$ and of the density is $\rho_{max}\sim8 \times10^{-4}$.}
\end{figure}

\section{Conclusion}\label{sec:dis}
Compact objects are well investigated both in GR and in several modified gravity theories, in $f(R)$ , Gauss-Bonnet and others \cite{JCAP 1501 },\cite{Phys.Rev. D93 (2016)}. In Ref.\cite{JCAP 1501 }it has been shown that in neutron stars, the dense matter in magnetic mean field, generated by magnetic fields of particles,can be underestood using a model with three meson fields and baryons octet. Consequently , the maximal mass of neutron stars is increased  and there is a possibility to have massive stars with $m>4 $. Furthermore we can have stable stars with high strangeness fraction .Furthermore in $f(R)$ or Gauss-Bonnet neutron stars , other branches of massive neutron stars are possible. In Ref. \cite{Phys.Rev. D93 (2016)}, the M-R profiles for static neutron star models obtained by the numerical solution of modified TOV equations in $f(R)$ gravity with different forms for $f(R)$,using perturbation theory. 
In this work  in the MG framework, the TOV equations for a static spherically symmetric spacetime, which corresponds to a compact star has been derived. A dynamical analysis showed that the system is locally unstable near the stationary point. The mass, pressure, density, metric function and scalar field were constructed numerically. Our results are in a good agreement with previous results in other types of modified gravity theories studied in  \cite{JCAP 1501} ,\cite{Phys.Rev. D93 (2016)}. An exact solution for a quark star  model with linear EoS was discovered. In comparison to $f(R)$ gravity case
\cite{ Phys.Lett. B742 (2015) 160}, we also found stable neutron star configurations. As an extended model, we solved TOV equations by assuming a polytropic EoS introduced in \cite{momeni/2015}. The qualitative behavior of the functions were studied. Our work exposes new features of MG as an alternative for GR and other types of modified gravity theories. It will be very interesting to study this TOV system of equation in other types of modified gravity theories like Mimetic $f(R)$ gravity \cite{Nojiri:2014zqa}.

\section*{Acknowledgments}
We would like to thank the referee for useful comments.

\end{document}